\documentclass[%
 reprint,
superscriptaddress,
 amsmath,amssymb,
 aps,
]{revtex4-2}

\usepackage{graphicx}
\usepackage{dcolumn}
\usepackage{bm}
\usepackage{color}

\usepackage{xcolor}
\usepackage{physics}
\usepackage{placeins}
\usepackage{caption}
\usepackage{lineno} 

\usepackage{subcaption}
\usepackage[font=tiny,labelfont=bf]{caption}
\usepackage{float}

\usepackage[colorlinks=true, allcolors=blue]{hyperref}

\makeatletter
\renewcommand{\fnum@figure}{\textbf{Figure \thefigure}}
\makeatother

\usepackage[font=small,labelfont=bf,
   justification=Justified,
   format=plain]{caption}

\begin{document}

\preprint{APS/123-QED}

\title{Effect of translational shear on interfacial structure in the viscous fingering instability}


\author{Zhaoning Liu}
\email{znliu@uchicago.edu}
\affiliation{Department of Physics and The James Franck and Enrico Fermi Institutes, University of Chicago, Chicago IL, 60637}
\author{Samar Alqatari}
\affiliation{Department of Physics and The James Franck and Enrico Fermi Institutes, University of Chicago, Chicago IL, 60637}
\author{Thomas E. Videb{\ae}k}
\affiliation{Department of Physics and The James Franck and Enrico Fermi Institutes, University of Chicago, Chicago IL, 60637}
\affiliation{Martin A. Fisher School of Physics, Brandeis University, Waltham MA, 02453}
\author{Sidney R. Nagel}
\affiliation{Department of Physics and The James Franck and Enrico Fermi Institutes, University of Chicago, Chicago IL, 60637}

\date{\today}

\begin{abstract}  
TEASER:  Oscillatory shear smooths fluid interfaces, stabilizing viscous fingering in miscible fluids. \\

We introduce applied shear as a method to control viscous fingering by smoothing the interface between miscible fluids. In the viscous fingering instability, a less viscous fluid displaces a more viscous one through the formation of fingers. 
The instability, which requires a confined geometry, is often studied in the thin gap of a quasi-two-dimensional Hele-Shaw cell. 
When the two fluids are miscible, the structures that form in the dimension traversing the gap are important for determining the instability onset.  We demonstrate with experiments and simulations that oscillatory translational shear of the confining plates changes the gap-averaged viscosity profile so that it become less abrupt at the finger tips. Increasing the amplitude or velocity of the shear delays the instability onset and decreases the finger growth rate. Shear can thus be used to stabilize a pair of miscible fluids against fingering. The results show a direct correlation between a smoother viscosity profile and delayed instability. 

\end{abstract}

\maketitle

\section{Introduction}

\begin{figure*}[t!]
    \centering
    \hypertarget{fig:demo}{}
    \includegraphics[width=\textwidth]{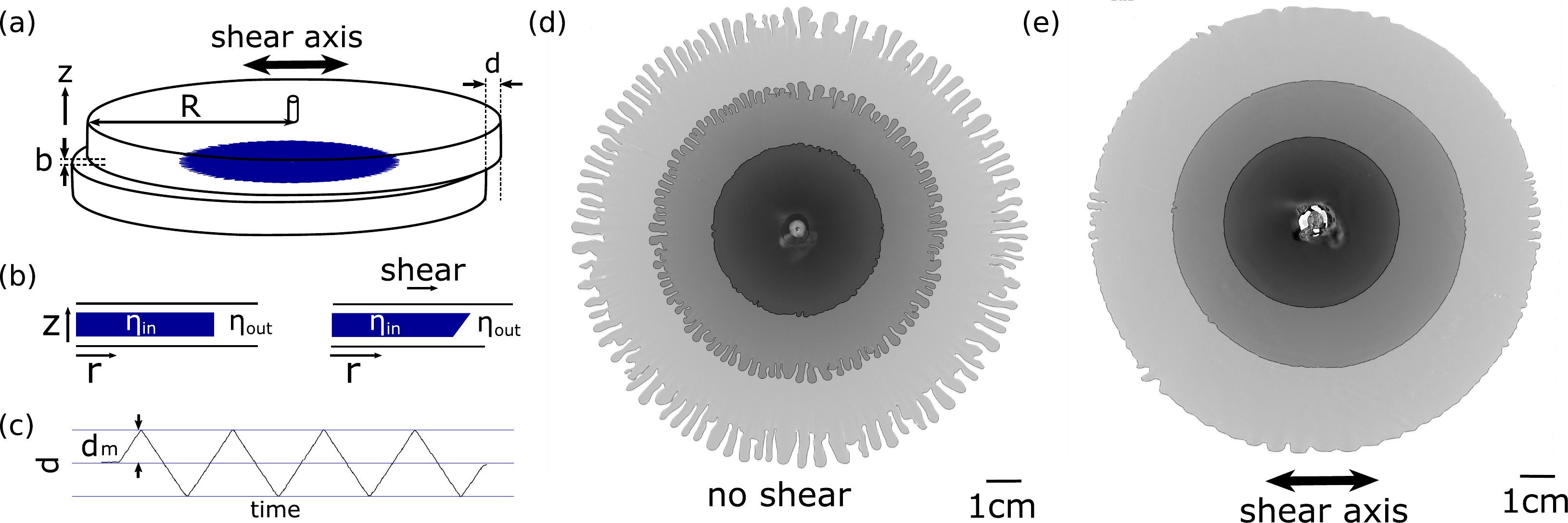}
    \caption{\textbf{Experiment: viscous fingering under oscillatory shear.} \textbf{(a)} Schematic for a radial Hele-Shaw cell where the top plate is displaced from the bottom one by an amount $d$. For measurements without shear, we align the plates with $d = 0$. The gap between the plates extends from $z=0$ to $z=b$. 
    $b$ is much smaller than the radius of the plates, $R$. Fluid is injected from an inlet at the center of the top plate.
    \textbf{(b)} A schematic of the effect of shear on a stationary vertical interface shown on left; after shear (right) the interface is tilted. 
    \textbf{(c)} The measured displacement, $d$, between the centers of the two plates is varied with a triangular waveform with amplitude $d_{\rm m}$ and velocity  $\pm V_{\rm s}$. \textbf{(d, e)} Comparison of interfaces from experiments (d) with no shear and (e) with shear with $d_{\rm m} = 3~\mathrm{mm}$ and $V_{\rm s} = 5.6~\mathrm{mm/s}$. Darker images show earlier times. Images with the same gray-scale in (d) and (e) are taken at the same radius. The smallest (\textit{i.e.}, darkest) image in (d) is taken when fingers first appear in the absence of shear.  In (e) at the same radius, the interface shows no finger formation. Fingers only begin to form in (e) at the second image. For both cases, $b=305~\rm{\mu m}$, $Q=133~\mathrm{\mu L/s}$, $\eta_{\rm out}= 218 \pm 5~\rm{mPa\cdot s}$, and $\eta_{\rm in}= 35 \pm 2~ \rm{m Pa\cdot s}$.
    }
    \label{demo}
\end{figure*}

Viscous fingering is a prototypical example of pattern formation. It occurs at the interface between two fluids when the fluid with lower viscosity displaces the other in a confined geometry such as a thin gap between two surfaces~\cite{saffman1958penetration}.  
At the instability onset, the inter-fluid interface develops lateral undulations that develop into fingers as the less-viscous fluid protrudes into the fluid that fills the rest of the gap. The conditions for this to occur generally depend on the viscosities of the two fluids, the interfacial tension, and the local velocity of the interface~\cite{saffman1958penetration,PhysRevFluids.8.113904}. 

As the interfacial tension drops, the fingers become thinner but do not become arbitrarily narrow 
, remaining finite even in the limit of miscible fluids where the interfacial tension nearly vanishes~\cite{paterson1985fingering, nagel2013new}. Paradoxically in that case, even though the stabilizing effect of the interfacial tension is removed, a regime occurs where the fluids become stable against lateral fingering; even though the invading fluid is less viscous, fingers may not form as they would for pairs of immiscible fluids. In that situation, for fingering to occur, the interface between the fluids must be sufficiently blunt in the direction spanning the gap (\textit{i.e.}, in the $z$-direction as illustrated in \hyperlink{fig:demo}{Fig.~\ref{demo}a,b})
~\cite{lajeunesse_3d_1997, lajeunesse_miscible_1999, bischofberger_fingering_2014,videbaek2020delayed,lister2024fingeringinstabilityselfsimilarradial}. 

The interface bluntness is controlled by the ratio of the inner to outer viscosity, $\eta_{\rm in}/\eta_{\rm out}$. In addition, if the injection rate is sufficiently low, the interface between miscible fluids can become smeared out due to interfluid diffusion; this leads to a delayed instability onset and a slower finger growth rate~\cite{PhysRevFluids.4.033902}. Thus, both the geometry  (\textit{i.e.}, the bluntness) and the diffusion independently alter the profile of the average viscosity near the finger tip. 
This raises issues of how sharp must the interface at the tip be in order for the fingering instability to occur and how does the variation of the gap-averaged viscosity near the tip control the onset and evolution of fingering. 
    
A common experimental platform for studying viscous fingering is in a radial Hele-Shaw cell, illustrated in \hyperlink{fig:demo}{Fig.~\ref{demo}a}, where two smooth horizontal glass plates of radius $R$ are separated by a gap of width $b$ with $b \ll R$~\cite{saffman1958penetration,saffman1986viscous,XU201692}. The gap is first filled with the fluid of viscosity $\eta_{\rm out}$ and then a lower viscosity fluid with viscosity $\eta_{\rm in}$ is injected through a small hole in the center of one of the plates.

For miscible fluids, the displacing fluid forms a thin tongue in the gap protruding between two layers of the displaced fluid. As  $\eta_{\rm in} / \eta_{\rm out} \rightarrow 1$, where the two fluids would have identical physical properties, the flow approaches a parabolic (Poiseuille) profile in the $z$-direction traversing the gap so that the gap-averaged viscosity profile, $\langle\eta(r)\rangle \equiv \frac{1}{b}\int_0^b \eta(z,r) dz$ gently increases as a function of radius, $r$.
As $\eta_{\rm in} / \eta_{\rm out}$ decreases, $\langle\eta(r)\rangle$ at the interface between the inner and outer fluids becomes more blunt at the tip of the invading fluid~\cite{lajeunesse_3d_1997, YORTSOS_SALIN_2006, bischofberger_fingering_2014}. 
The quasi-two-dimensional lateral pattern becomes unstable to fingering for $\eta_{\rm in}/\eta_{\rm out} \lesssim 0.3$. This transition was associated with shock formation in $\langle\eta(r)\rangle$~\cite{lajeunesse_3d_1997, lajeunesse_miscible_1999}. 
So far, it has not been possible to isolate the specific role of interface shape from other physical properties  such as mobility, viscosity ratio or diffusion. A recent theoretical study corroborated that a discontinuity in $\langle\eta(r)\rangle$ at the interface tip is required for fingering onset~\cite{lister2024fingeringinstabilityselfsimilarradial}. It has remained unclear whether the onset can be continuously delayed by progressively smoothing the interface.  

In this paper, we use translational oscillatory shear between the parallel plates of the Hele-Shaw cell to perturb the interface actively. This allows the shape of the interface to be altered independently from either the viscosity ratio or diffusion and directly tests whether the interface shape is indeed the critical driver of the instability. This experiment quantitatively correlates smoothness to the instability.

If the two fluids were initially stationary with the tip of their interface perfectly vertical, then shearing the top plate would tilt the interface as shown in the schematic in \hyperlink{fig:demo}{Fig.~\ref{demo}b}. 
In this simplified view with no shear, the gap-averaged viscosity would have a discontinuous jump at the interface. However, when shear is applied, the tilting of the interface causes the gap-averaged viscosity to increase smoothly from the inner to the outer fluid. When the fluids are not at rest, the contour of the inter-fluid interface will depend on the shear-displacement amplitude, $d_{\rm m}$, and the relative velocity of shear, $\pm V_{\rm s}$, with respect to the advancing interface, $U$, which is proportional to the fluid injection rate, $Q$. Thus shear can present an independent way to alter $d\langle\eta(r)\rangle/dr$.

Because the cell is circular and the shear is along one chosen axis, the direction of shear relative to finger propagation varies around the pattern. Thus, we observe the effects of shear in two orthogonal directions.  In this paper, we focus on fingers growing in the parallel direction where shear delays the instability onset and lowers the growth rate of fingers once formed. The effect of shear on fingers in the perpendicular direction will be presented in a later paper. 


\section{Results}
\subsection{Experimental platform}

Our Hele-Shaw cell consists of two large, flat, circular glass plates with a radius $R = 14$ cm as illustrated in \hyperlink{fig:demo}{Fig.~\ref{demo}a}.
The gap spacing between the plates was kept uniform at $b=305~\rm \mu m$ by using six spacers of equal height placed around the cell edge. The top plate is maintained in contact with the washers by
weights equally placed around the perimeter. The fluids are injected through a hole in the center of the top plate. Fingering patterns are imaged from below the plates.  

During the fluid injection, the plates are cyclically sheared with respect to each other along one axis with a triangular waveform to a constant maximum amplitude, $d_\mathrm{m}$ as shown in \hyperlink{fig:demo}{Fig.~\ref{demo}c}. In different experiments, we varied $d_\mathrm{m}$ and shear speed, $V_{\rm s}$:  $0.3~{\rm mm} < d_{\rm m} < 6~{\rm mm}$ and $1.4~{\rm mm/s} < V_{\rm s} < 28~{\rm mm/s}$.  

Both the inner and outer fluids are mixtures of glycerol and water whose viscosities are $\eta_{\rm in}= 35 \pm 2~\rm mPa\cdot s$ and $\eta_{\rm out}= 218 \pm 5~\rm mPa\cdot s$ respectively.  The inner fluid is dyed, so that the concentration profile of the inner fluid, $C(r)$, can be measured from the transmitted light intensity as discussed in \hyperlink{Sec:Methods}{Methods}~\cite{bischofberger_fingering_2014,PhysRevFluids.4.033902}.
Initially, the gap between the plates is fully filled with the more viscous fluid. The less viscous displacing fluid is then injected using a syringe pump at a constant volume rate,  $Q$, with values between $67~\mathrm{\mu L/s} \leq Q \leq 533~\mathrm {\mu L/s}$.  This ensures that the Péclet number is large enough, $\mathrm{Pe} \equiv U b/D \gg 10^3$, where $D$ is the effective interfluid diffusivity, that there is a well-defined interfluid interface~\cite{PhysRevFluids.4.033902} such that $\langle\eta(r)\rangle = (\eta_{\rm in}-\eta_{\rm out})\cdot C(r)+\eta_{\rm out}$.  Details of the experiment are given in \hyperlink{Sec:Methods}{Methods}.    


Our experiments investigate how the application of shear between the confining plates changes the onset radius of the fingering instability, $R_{\rm on}$, and the growth rate of the fingers once they are formed, $\Gamma$.  We measure the concentration profile of the inner fluid, $C(r)$, and correlate its shape with $R_{\rm on}$. We focus on fingers growing in the direction parallel to the shear axis. 

\subsection{Effect of shear on onset radius and finger growth rate} 

\hyperlink{fig:demo}{Figure~\ref{demo}d} shows three superimposed images from a conventional experiment without shear between the plates. The darkest image at the center, taken at the earliest-time, shows fingers just becoming visible; the lighter patterns, taken at later times, show well-developed fingers emerging in all directions.  Two relevant features of these patterns are the onset of finger growth at a radius, $R_{\rm on}$, which is set (in both miscible and immiscible pairs of fluids ) by the viscosity ratio, $\eta_{\rm in} / \eta_{\rm out}$~\cite{bischofberger2015island}, and a wavelength characterizing the finger widths.

\hyperlink{fig:demo}{Figure~\ref{demo}e} shows the corresponding images taken while the plates were cyclically sheared during fluid injection. The shear is along the axis of the double-headed arrow.  Aside from the shear, the two experiments had the same fluids, gap spacing, and injection rate.
In \hyperlink{fig:demo}{Fig.~\ref{demo}e}, the innermost pattern is smooth with no indication of incipient finger growth. This should be compared to the image with the same gray scale in \hyperlink{fig:demo}{Fig.~\ref{demo}d} taken at the same radius  where fingers had already formed in the absence of shear.  Fingers only become visible at larger radii as seen in the image with a radius approximately $1.6$ times larger than the first. Shear delays the onset of fingering substantially.   

\begin{figure}[t!]
\begin{center}
\hypertarget{fig:onsets}{}
\includegraphics[width=\linewidth]{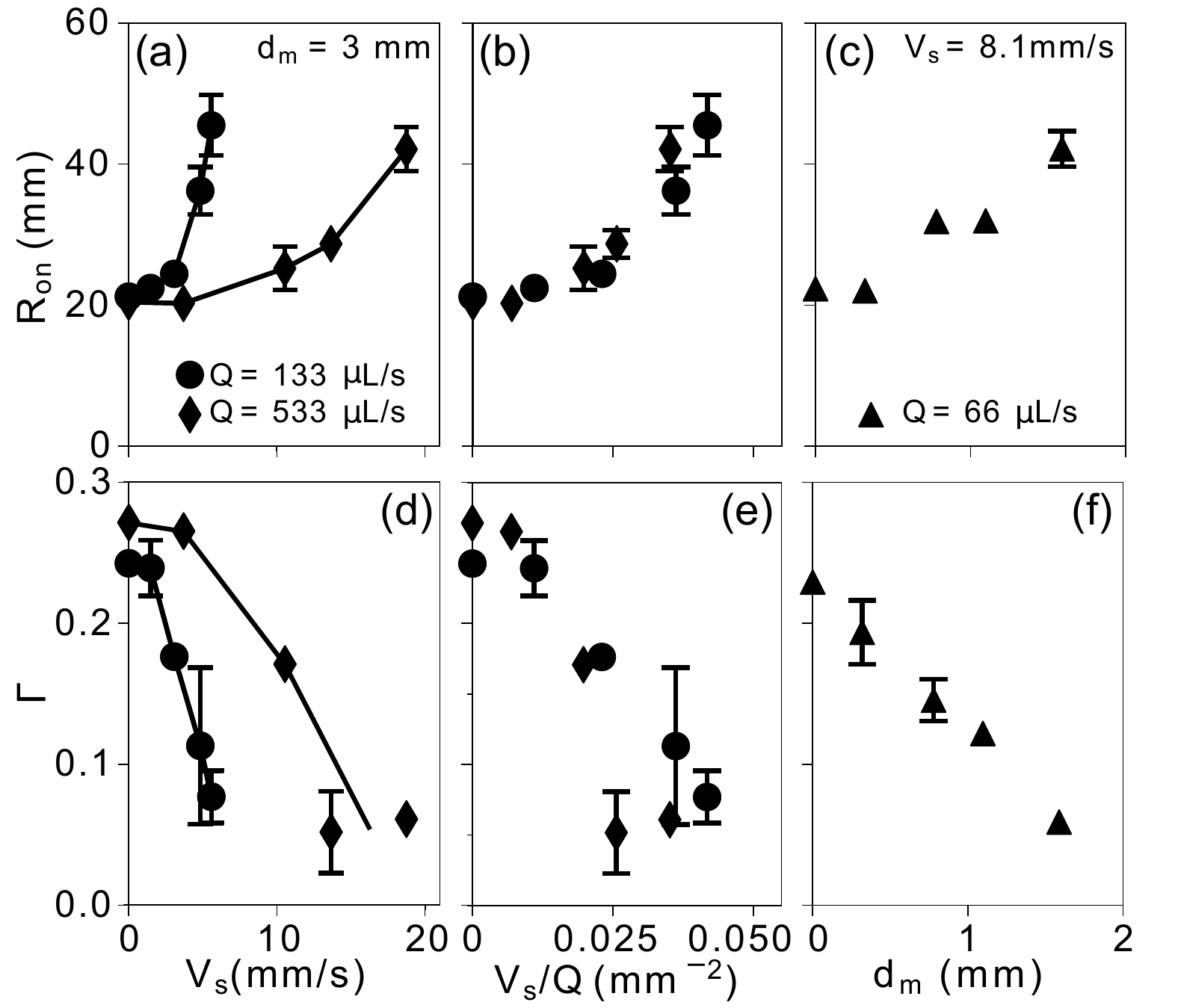}
\caption{\textbf{Effect of shear on instability onset and growth for fingers parallel to the shear axis.} \textbf{(a-c)} show data for onset radius and \textbf{(d-f)} show data for growth rate.  \textbf{(a,d)} Onset radius, $R_{\rm on}$, and growth rate, $\Gamma$ versus shear speed $V_{\rm s}$. 
\textbf{(b,e)} Same data as in \textbf{(a,d)}, but $V_{\rm s}$  is normalized by the injection rate, $V_{\rm s}/Q$. The curves collapse for the different $Q$.  \textbf{(c,f)} $R_{\rm on}$ and $\Gamma$ versus $d_{\rm m}$. For all data shown in this paper, the error bars, unless explicitly shown, are comparable to or smaller than the size of the symbols.
}
\label{onsets}

\end{center}
\end{figure}

\hyperlink{fig:onsets}{Figure~\ref{onsets}} quantifies how increasing the shear speed, $V_{\rm s}$, or shear amplitude, $d_{\rm m}$, delays the instability onset.  As shown in \hyperlink{fig:onsets}{Fig.~\ref{onsets}a}, $R_{\rm on}$ increases with increasing $V_{\rm s}$ while $d_{\rm m}$ is held constant. This delay persists across different injection rates, $Q$.  Normalizing the shear speed by the injection rate, $V_{\rm s}/Q$, scales the data onto a single curve as shown in \hyperlink{fig:onsets}{Fig.~\ref{onsets}b}.  This suggests that the delay in onset is governed by the shear speed relative to the speed of the advancing interface.
\hyperlink{fig:onsets}{Figure~\ref{onsets}c} shows that increasing the shear amplitude $d_{\rm m}$ while holding $V_{\rm s}$ and $Q$ constant also increases  $R_{\rm on}$. 

Once the fingers have formed, shear also reduces the finger growth rate: $\Gamma \equiv dR_{\mathrm{f}}/dR_{\mathrm{tip}}$, where $R_{\mathrm{f}}$ is the finger length, and $R_{\mathrm{tip}}$ is the outer extent of the finger. Similar to the case without shear~\cite{videbaek2020delayed}, soon after onset, driven by pressure gradients in the bulk of the fluids~\cite{gowen_2024}, $R_{\mathrm{f}}$ varies linearly with $R_{\mathrm{tip}}$, as shown in \hyperlink{Sec:SI}{SI}. \hyperlink{fig:onets}{Figure~\ref{onsets}d-f} shows $\Gamma$ just after onset, before $R_{\mathrm{tip}}-R_{\mathrm{on}}$ reaches 10 mm. 
$\Gamma$ decreases with increasing $V_\mathrm{s}$ as shown in \hyperlink{fig:onets}{Fig.~\ref{onsets}d}; as for $R_{\rm on}$, the data for different injection rates collapse onto one curve when $\Gamma$ is plotted versus $V_\mathrm{s}/Q$ as shown in \hyperlink{fig:onets}{Fig.~\ref{onsets}e}. Likewise, increasing $d_{\mathrm{m}}$ at fixed $V_\mathrm {s}$ and $Q$ also suppresses $\Gamma$ as shown in (\hyperlink{fig:onets}{Fig.~\ref{onsets}f}). These results demonstrate that increasing either shear amplitude or speed increases the radius of onset, $R_{\rm on}$, and also slows down the subsequent finger-growth dynamics, $\Gamma$.

\subsection{Transverse structure across the gap}

\begin{figure}[t!]
\begin{center}
\hypertarget{fig:concentrationprofile}{}
\includegraphics[width=7.6cm]{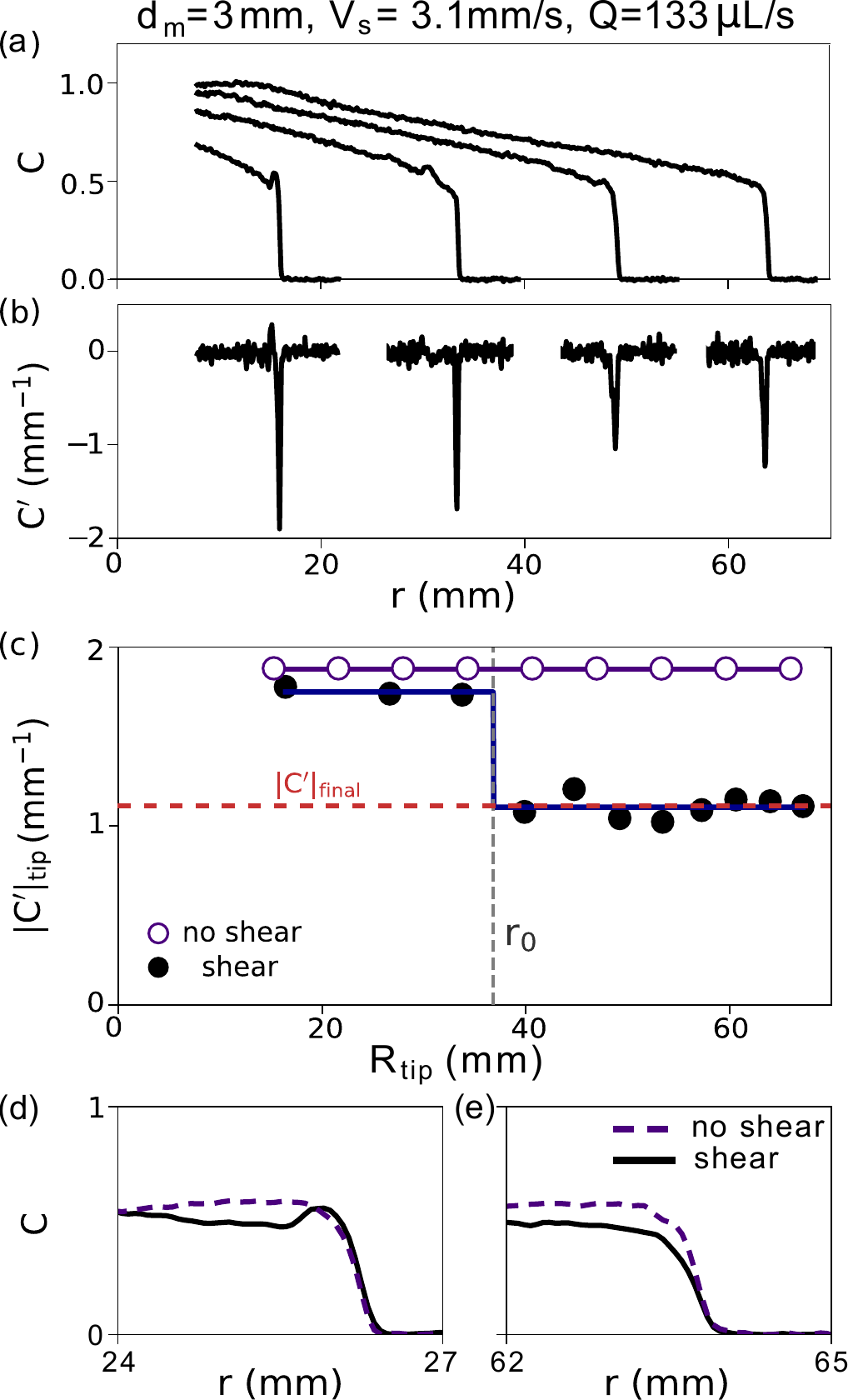}
\caption{\label{concentrationprofile}
\textbf{Effect of shear on the interface concentration profile.} 
\textbf{(a)} Selected profiles of inner-fluid concentration, $C(r)$ taken at four positions, from left to right, $R_\mathrm{tip}\approx16~\mathrm{mm},33~\mathrm{mm},49~\mathrm{mm},\mathrm{and}~64~\mathrm{mm}$.  \textbf{(b)} Radial derivative of the concentration profile, $C'(r)$, measured near the finger tip at the same positions as in \textbf{(a)}. 
\textbf{(c)} $|C'|_{\rm tip}$ versus the position of the tip $R_\mathrm{tip}$. For the case with shear the fit is to the step-function described in the text. Without shear, $|C'|_{\rm tip}$ is approximately constant. \textbf{(d)} At $R_\mathrm{tip}\approx 26~\mathrm{mm}<r_\mathrm{0}\approx 37~\mathrm{mm}$, the profiles with and without shear are similarly blunt. \textbf{(e)} At $R_\mathrm{tip}\approx 64~\mathrm{mm}>r_\mathrm{0}$, the profile with shear is smoother.
}
\

\end{center}
\end{figure}

We now investigate the correlation between the interface profile in the $z$-direction across the gap and the onset of the fingering. To do this, we measure the gap-averaged inner-fluid concentration profile $C(r)$ along the radial direction $r$ from the center of the inner fluid to the interface along a finger, as described in \hyperlink{Sec:Methods}{Methods}. 
By measuring the concentration profile during the course of an experiment, we can measure the interface evolution, as shown in \hyperlink{fig:concentrationprofile}{Fig.~\ref{concentrationprofile}a}.

To measure the smoothness or bluntness of the interface we take the derivative of the concentration profile $C'(r) \equiv dC(r)/dr$. As shown in \hyperlink{fig:concentrationprofile}{Fig.~\ref{concentrationprofile}b}, $C'(r)$ dips abruptly at the end of a finger. The maximum of its absolute value, $|C'|_{\rm tip}$, reflects how blunt the finger is at its tip.  The peaks in $|C'(r)|$ decrease as the interface expands while being sheared.  The corresponding evolution of the concentration profiles without shear is shown in the \hyperlink{Sec:SI}{SI}.

\hyperlink{fig:concentrationprofile}{Figure~\ref{concentrationprofile}c} shows that when there is shear, $|C'|_{\rm tip}$ remains constant until the interface reaches $r_0 \approx 37~\rm mm$, at which point it rapidly drops to a lower plateau value $|C'|_{\rm tip} = |C'|_{\rm{final}}$. 
This behavior defines a  length $r_0$, the \textit{smoothing radius}, where shear begins to modify the interface. 
We quantify this transition with a step function: 
$$|C'|_{\rm tip} = |C'|_{\rm{final}} + \Delta C' \Theta(r_0 - r),$$ 
where $ \Theta(...)$ is the Heaviside step function.  When there is no shear, $|C'|_{\rm tip}$ remains constant as the inner fluid expands.

\hyperlink{fig:concentrationprofile}{Figures~\ref{concentrationprofile}d,e} show the concentration profiles near the tip for the case with shear compared to the case when there is no shear. When $R_\mathrm{tip} < r_0$ the profile shapes are nearly the same whereas when $R_\mathrm{tip} > r_0$, the profile for the sheared case is smoother.

\begin{figure}[t!]
\hypertarget{fig:speedcompetition}{}
\includegraphics[width=\linewidth]{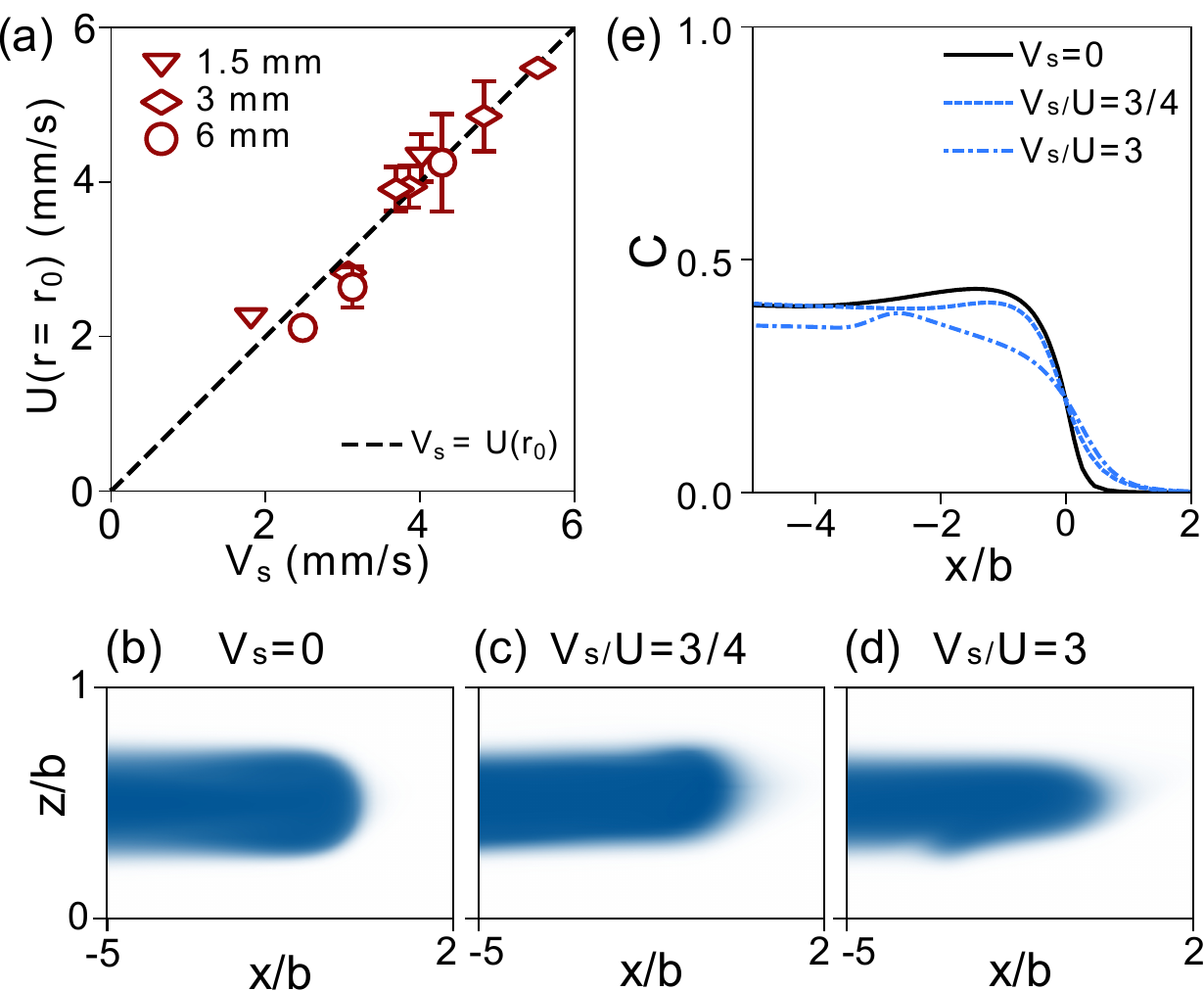}
\caption{\label{speedcompetition} 
\textbf{Competition between shear speed and interface velocity.}
\textbf{(a)} The experimentally measured interface speed at $r=r_0$, $U(r=r_0)$, versus the shear speed, $V_{\rm s}$. The dashed line, with slope one, shows that $r_0$ is determined by when $U(r=r_0) = V_{\rm s}$. 
\textbf{(b-d)} Two-dimensional COMSOL simulation results showing the shape of the inner fluid (blue) around its tip at three shear speeds: \textbf{(b)} $V_{\rm s} = 0$, \textbf{(c)} $V_{\rm s} / U = 3/4$, and \textbf{(d)} $V_{\rm s} / U = 3$. 
\textbf{(e)} Concentration profiles from the simulations shown in \textbf{(b-d)}. Compared to the case of no shear, the curve for $V_{\rm s} < U$ is only slightly perturbed whereas for $V_{\rm s} > U$, the tip is thinner and smoother.
}
\end{figure}

In order for the applied shear to alter  $C(r)$ appreciably, one would expect that the shear speed $V_{\rm s}$ would need to be comparable to (or larger) than the interface speed, $U$. If it were much smaller, it would not perturb the flow appreciably. At fixed injection rate, because of the radial geometry of the cell, $U \propto 1/r$ so that $U(r)$ eventually decreases below $V_{\rm s}$.  The data of \hyperlink{fig:concentrationprofile}{Fig.~\ref{concentrationprofile}c} shows that the thinning of the profile occurs abruptly at $r = r_0$. Therefore, in \hyperlink{fig:speedcompetition}{Fig.~\ref{speedcompetition}a}, we compare the measured interface speed at $r = r_0$, $U(r=r_0)$, with $V_{\rm s}$. The dashed line indicates where the two speeds are equal.  The data is consistent with $r_0$ being determined by where the interface speed becomes comparable to $V_{\rm s}$.

To understand how the competition between $V_\mathrm{s}$ and $U$ affects the interface profile, we simulate  the flow within the gap using COMSOL. The specifics and parameters used are described in the \hyperlink{Sec:Methods}{Methods} section. In order to isolate how the steady-state fluid interface depends on the relative shear and injection velocities, we use a linear cell so that $U$ is constant throughout each simulation. While the results are thus not directly comparable with experiments in a circular geometry where the interface velocity varies inversely with radius, they provide physical insight into the underlying smoothing mechanism.

The three profiles shown in \hyperlink{fig:speedcompetition}{Fig.~\ref{speedcompetition}(b-d)} show how shear affects the interface shape near its tip. 
\hyperlink{fig:speedcompetition}{Figure~\ref{speedcompetition}b} shows the profile when there is no shear, \textit{i.e.}, $V_{\rm s}=0$ while \hyperlink{fig:speedcompetition}{Fig.~\ref{speedcompetition}c,d} show the profiles for $V_{\rm s}/ U = 3/4$ and $V_{\rm s} / U = 3$ respectively. 
\hyperlink{fig:speedcompetition}{Figure~\ref{speedcompetition}e} shows the concentration profiles, $C(r)$, for these three cases.  When $V_{\rm s}< U$ the thickness of the intruding tongue of inner fluid is approximately the same as when there was no shear.  When $V_{\rm s} > U$, the tongue becomes appreciably narrower and has a more tapered tip.  These simulations are consistent with our experiments and the argument that the smoothing occurs appreciably only when the applied shear is at least as large as the interfacial velocity.

In the \hyperlink{Sec:SI}{SI}, we also use these simulations to show how the inner-fluid profile becomes established. The thickness of the tongue continues to decrease slightly during the first few oscillations of shear.

\subsection{Onset and bluntness correlation}

\begin{figure}[t!]
\hypertarget{fig:r0_Ron}{}
\includegraphics[width=\linewidth]{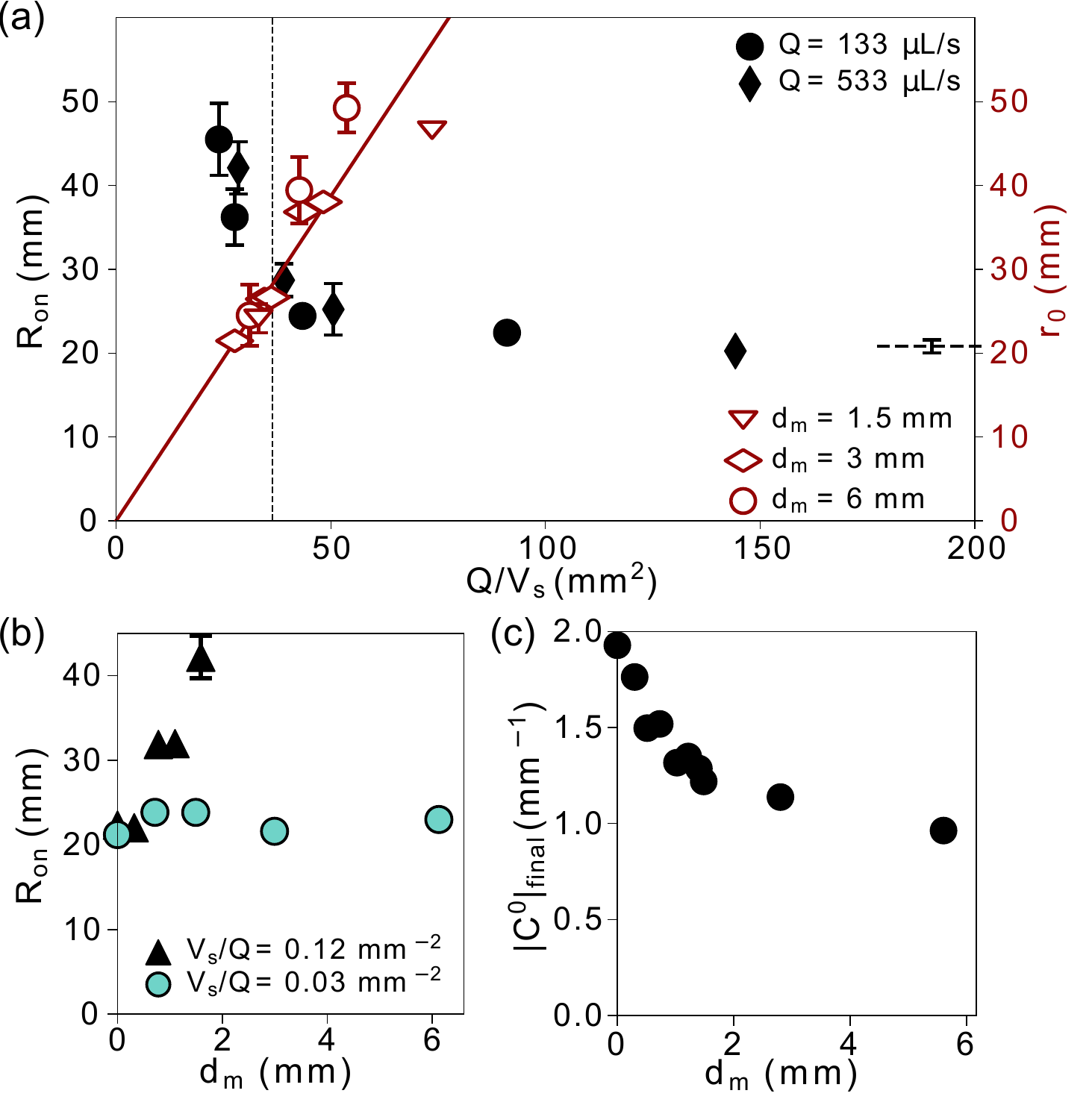}
\caption{\label{r0_Ron} 
\textbf{Competition between fingering onset, $R_{\rm on}$, and smoothing radius, $r_0$, under shear.}
\textbf{(a)} $r_0 \propto Q/V_{\rm s}$ (red) while $R_{\rm on}$ (black) decreases until the data sets cross at $R_{\rm on} \approx r_0$ shown by vertical dashed line. To the left of their intersection, $r_0 < R_{\rm on}$ resulting in a significant delay in $R_{\rm on}$. To the right, 
$r_0 > R_{\rm on}$ so the smoothing has only a mild influence on $R_{\rm on}$. 
(Same $R_\mathrm{on}$ data set as in \protect\hyperlink{fig:speedcompetition}{Fig.~\ref{onsets}b}.)
\textbf{(b)} $R_{\rm on}$ (black) increases significantly with shear amplitude, $d_{\rm m}$, when $V_{\rm s}/Q > 0.027~\mathrm{mm}^{-2}$, the intersection value found in (a). For $V_{\rm s}/Q < 0.027~\mathrm{mm}^{-2}$, $R_{\rm on}$ (teal) remains nearly independent of $d_{\rm m}$. 
\textbf{(c)} $|C'|_{\rm final}$ decreases with $d_{\rm m}$. For $d_\mathrm{m}=1.5~\mathrm{mm},~3~\mathrm{mm}, \mathrm{and}~6~\mathrm{mm}$, $|C'|_\mathrm{final}$ is averaged over different values of $V_\mathrm{s}/Q$ as those in \protect\hyperlink{fig:speedcompetition}{Fig.~\ref{Cf_Ron}a}. For all the other values of $d_\mathrm{m}$, $V_\mathrm{s}/Q = 0.12\pm0.01~\mathrm{mm^{-2}}$.
}
\end{figure}

\begin{figure}[t!]
\centering
\hypertarget{fig:Cf_Ron}{}
\includegraphics[width=\linewidth]{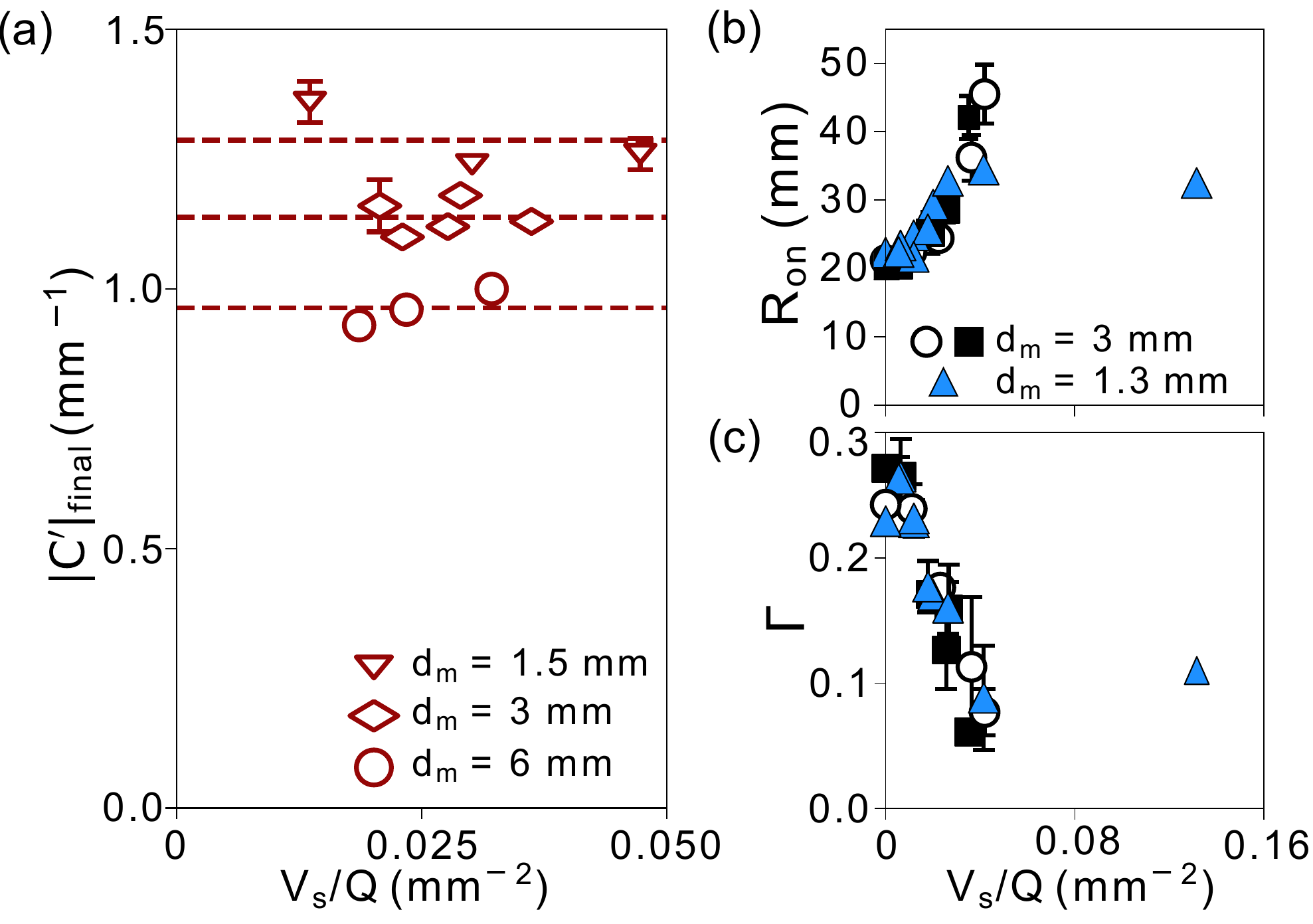}
\caption{\label{Cf_Ron} 
\textbf{Role of shear-induced interface smoothness on instability suppression.}  
\textbf{(a)} Final value of profile bluntness, $|C'|_{\rm final}$, is insensitive to normalized shear speed, $V_{\rm s}/Q$, for different shear amplitudes, $d_\mathrm{m}$.  At $d_{\rm m}=1.3~ \mathrm{mm}$, $R_{\rm on}$ \textbf{(b)} and $\Gamma$ \textbf{(c)} plateau at high $V_{\rm s}/Q$, showing a limit set by the corresponding $|C'|_{\rm final}$. In comparison, at $d_\mathrm{m}=3~\mathrm{mm}$ (same data as in \protect\hyperlink{fig:speedcompetition}{Fig.~\ref{onsets}b}), where $|C'|_{\rm final}$ is smaller than that of $d_\mathrm{m}=1.3 ~\mathrm{mm}$, $R_{\rm on}$ and $\Gamma$ extend beyond  the measurement limit.
}
\end{figure}

Having identified $r_0$ as the characteristic radius where smoothing due to shear begins, we now examine how it relates to the onset of the fingering instability. There are two important parameters, $V_{\rm s}/Q$ and $d_{\rm m}$, for determining how shear smooths the interface.  
We show here that in order to affect the fingering onset: (i) the shear speed, $V_{\rm s}/Q$, must be large enough that the smoothing occurs prior to the onset of finger growth, \textit{i.e.}, $r_0 < R_{\rm on}$; and (ii) the amplitude of shear, $d_{\rm m}$, which alters the degree to which the profile becomes smoother, can likewise delay the instability only when  $r_0 < R_{\rm on}$.

\textit{Role of shear speed}:  If the interface smoothing occurs \textit{after} onset (\textit{i.e.,} $r_0>R_{\rm on}$), there is no opportunity for shear to delay the onset and one would expect $R_{\rm on}$ to remain nearly unchanged.  
This is corroborated by the data for $R_{\rm on}$ and $r_0$ versus $Q/V_{\rm s}$ in \hyperlink{fig:r0_Ron}{Fig.~\ref{r0_Ron}a}. 
The plot shows $r_0 \propto Q/V_{\rm s}$ while $R_{\rm on}$ does not vary dramatically until the two data sets cross, (\textit{i.e.}, $r_0 \approx R_{\rm on}$) shown by the vertical dashed line. 
Only to the left of this line, where the smoothing occurs before onset, does $R_{\rm on}$ increase significantly.

\textit{Role of shear amplitude}: The relative values of $R_{\rm on}$ and $r_0$ also determines the effect of shear amplitude, $d_{\rm m}$ on the onset radius $R_{\rm on}$.  As shown in \hyperlink{fig:r0_Ron}{Fig.~\ref{r0_Ron}b}, at $V_{\rm s}/Q = 0.12~\rm mm^{-2}$ (black data), where $r_0 < R_{on}$, increasing $d_{\rm m}$ markedly delays $R_{\rm on}$.  \hyperlink{fig:r0_Ron}{Figure~\ref{r0_Ron}c} shows that $|C'|_{\rm{final}}$ decreases with increasing $d_{\rm m}$, indicating that the shear amplitude controls the final profile smoothness. As a result, the delayed onset with increasing $d_{\rm m}$ at high shear speeds also correlates with a smoother profile. 

In contrast, when  $V_{\rm s}/Q $ decreases to $0.03~\rm mm^{-2}$, where $r_0 > R_{on}$, increasing $d_{\rm m}$ to $6~\rm mm$ has little effect on $R_{\rm on}$.  This corroborates that unless smoothing starts prior to onset, the shear amplitude, which controls how much smoothing occurs, does not affect the instability. 


\hyperlink{fig:Cf_Ron}{Figure~\ref{Cf_Ron}a} shows that $|C'|_{\rm{final}}$ does not vary with $V_{\rm s}/Q$, but only depends on $d_{\rm m}$. 
Thus, once $r > r_0$, the profile stays a constant smoothness even as $V_{\rm s}/Q$ increases. That is, $|C'|_{\rm{tip}}$ does not decrease below $|C'|_{\rm{final}}$ which is the lower limit (depending only on the shear amplitude but not shear speed) for the smoothness.

Consistent with the idea that the smoothness of the interface controls the onset of the fingering instability and the subsequent finger growth, this lower limit in the profile smoothness also sets the limiting value for both both $R_{\rm on}$ and $\Gamma$.  This is seen in  \hyperlink{fig:Cf_Ron}{Fig.~\ref{Cf_Ron}b,c} where at $d_{\rm m} = 1.3~\rm mm$, both $R_{\rm on}$ and $\Gamma$ reach a plateau for $V_{\rm s}/Q> 0.04~ \rm{mm^{-2}}$ that is visible within our measurement window. As the shear amplitude increases to $d_{\rm m} = 3~\rm mm$, both $R_{\rm on}$ and $\Gamma$ continue to evolve until they exceed our measurement range.

\section{Discussion and Conclusion}

Previous experiments have indicated that a necessary condition for the formation of a viscous fingering instability in miscible fluids is the existence of a sufficiently blunt interface in the $z$-direction spanning the gap~\cite{lajeunesse_3d_1997, lajeunesse_miscible_1999, bischofberger_fingering_2014,videbaek2020delayed,lister2024fingeringinstabilityselfsimilarradial} and 
here we have directly addressed the role of the viscosity contrast in the case of low diffusion by applying oscillatory shear to perturb the three-dimensional interface profile. This work confirms the fundamental correlation between the smoothness of the gap profile and the onset of the instability. It demonstrates the conditions under which shear produces a smoother interface and how the fingering onset radius and growth rate are influenced by the interface smoothness. There are two important shear variables: the shear speed and the shear amplitude that play distinctly different roles. 

\subsection{Different roles of shear speed,  $V_{\rm s}$ and shear amplitude, $d_{\rm m}$}

In order for shear to distort the inner-fluid profile significantly, $V_{\rm s}$ must be comparable to the  velocity of the interface $U(r)$ as shown in \hyperlink{fig:speedcompetition}{Fig.~\ref{speedcompetition}a}. If $V_{\rm s}$ were much smaller than this, the dynamics would naturally be dominated by the injection rate, $Q$, rather than by the shear.  Our data indicate that it is only in the regime $V_{\rm s} > U(r)$ that the shear can effectively smooth the interface, $C(r)$.

In a circular cell, as in our experiments, $U(r) \propto Q/r$.  This allows the condition $V_{\rm s} = U(r)$ to occur at different radii. Initially, at small $r$, the interface moves rapidly compared to the shear speed, $V_\mathrm{s}$, of the plates; at later times, after the interface has expanded, the interface speed drops below $V_{\rm s}$.  
Thus, at large enough radius, the condition stated above for smoothing will always be met at some radius $r = r_0$. The competition between $V_{\rm s}$ and $U(r)$ explains why $r_0 \propto (Q/V_{\rm s})$ in \hyperlink{fig:r0_Ron}{Fig.~\ref{r0_Ron}a}.  The speed of interface growth is $U(r=r_0)\approx Q/(2\pi b r_0) =V_{\rm s}$, so that $r_0\approx Q/(2\pi b V_{\rm s})$. 
For $r < r_\mathrm{0}$, $|C'|_{\rm tip}$,  remains constant and then drops abruptly (\textit{i.e.}, within our resolution) to $|C'|_{\rm final}$ at $r_\mathrm{0}$ where the interface becomes smoother.

This alteration of the structure in the $z$-direction changes the fingering instability onset and the subsequent growth rate. Crucially, our experiments demonstrate that the delayed onset can only occur when the effect of shear occurs early enough that it precedes the onset. 
 
The shear amplitude, $d_{\rm m}$, can only appreciably affect the inner fluid profile after the smoothing has started to occur, that is after $r \ge r_0$. The interplay between $R_{\rm on}$ and $r_0$ reveals a transition in the instability dynamics and growth rate, $\Gamma$. When $R_{\rm on} \le r_0$ the dynamics of the instability proceeds as if there were no shear-induced thinning of the profile.  In this region the fingering dynamics is insensitive to $r_0$. This remains true even at high amplitudes, $d_{\rm m}$. However, when $R_{\rm on} > r_0$ the applied shear not only delays $R_{\rm on }$ but also reduces the subsequent finger growth rate $\Gamma$.  Consistent with the trend of $R_{\rm on}$, $\Gamma$ decreases with increasing $V_{\rm s}$ and $d_{\rm m}$.

One can revert to the no-shear limit by allowing either $d_{\rm m}\rightarrow0$ or $V_{\rm s}\rightarrow0$. As $d_{\rm m}\rightarrow0$, $|C'|_{\rm{final}}\rightarrow |C'|_{\rm{no~shear}}$, meaning there is no reduction in bluntness and therefore no suppression of $R_{\rm on}$ or $\Gamma$. On the other hand, as $V_{\rm s}\rightarrow0$, $r_0\rightarrow\infty$ , the interface never reaches the radius where $|C'|_{\rm tip}$ begins to drop. Hence,  the smoothing effect cannot emerge in finite time, and no instability suppression occurs. Thus, while both the parameters $d_{\rm m}$ and $V_{\rm s}$ cause a delay in onset, they do so in different ways.

\subsection{Shear as a control parameter}
This work introduces applied shear as a relevant perturbation which can be used to control viscous fingering. It offers practical advantages since it allows additional control over instability suppression independent of intrinsic fluid properties. 

What is not intuitive to us is how the shape of the profile, $C(r)$, gets established. This is observed in our experimental measurements of the concentration profiles (\hyperlink{fig:concentrationprofile}{Fig.~\ref{concentrationprofile}}) and corroborated in the COMSOL simulations, which are discussed in the \hyperlink{Sec:SI}{SI}. Initially, as the injection starts and the shear is applied,  the shape of the profile near the tip varies dramatically.  After only a few cycles, it reaches a condition where the local profile returns to nearly the same shape on each oscillation. This is seen clearly in \hyperlink{fig:concentrationprofile}{Fig.~\ref{concentrationprofile}c} where $C'(r)$ reaches a constant value $|C'|_{\rm final}$. Thus, the interplay of the dynamics that creates the shape of the inner fluid tongue needs further study. 

Other attempts to control the instability have been attempted.  By lowering the injection rate, 
diffusion effects were increased to blur the concentration profile. While this initially delayed the instability, a different fingering instability, with different wavelength and growth rate, emerged when the Péclet number was sufficiently low ~\cite{PhysRevFluids.4.033902}.
Other studies have used non-uniform~\cite{al2012control}, or time-dependent~\cite{PhysRevLett.115.174501} gap geometries, variable injection rates~\cite{PhysRevLett.102.174501,PhysRevE.92.041003}, deformable membranes~\cite{PhysRevLett.108.074502} and electric fields~\cite{gao2019active} to modify the instability.  Each of these methods has drawbacks as well as advantages. Applying mechanical shear offers a robust alternative including the possibility of temporal control. Similarly, additive manufacturing and biomedical engineering could adopt shear-driven smoothing to enhance coating uniformity or vascular network fabrication.

\hypertarget{Sec:Methods}{}
\section{Methods}
\textit{Experiments}: 
The Hele-Shaw cell consisted of two circular confining glass plates with radii $R=14$ cm and thickness greater than 1.27 cm in order to prevent bending.  The gap spacing between the plates was $b=305~\rm \mu m$ and was kept uniform by six spacers located around the periphery of the plates. The fluids were injected using a syringe pump (NE-1000 from New Era Pump Systems Inc.) via a hole in the top plate of diameter 1.6 mm.  The two plates were carefully aligned and the outer fluid was injected to the edge of the cell. Before injecting the inner fluid, the outer fluid residue and  bubbles inside the inlet were removed by flushing the less-viscous displacing inner fluid through the injection tube to the inlet and pumping it out from the waste tube. The valve for the waste tube was then closed. The fluid viscosities were measured by an Anton Paar MCR 301 rheometer.  

The bottom plate had no hole so that during the applied shear there would not be any perturbation to the flow caused by an opposing hole. This plate was leveled on top of three T-slotted framing rails with bumpers between the rails and  plate. Screws were mounted on the rails to secure the bottom plate at rest. Bumpers were used between screws and the bottom plate to reduce vibration. The top plate was enclosed by an aluminum ring connected to a 2-inch Feedback Rod Linear Actuator (FA-PO-35-12-2 from Firgelli Automations). An Arduino with a High Power Motor Drive (HiLetgo BTS7960 43A) was programmed to drive the actuator back and forth with a constant speed and amplitude.   

Using two parallel lines, one on each plate, perpendicular to the direction of shear, the relative plate displacement was measured (\hyperlink{fig:demo}{Fig.~\ref{demo}c}).  Considering the maximum displacements $d_\mathrm{m}$ were not always captured by the camera with finite frame rate, we measure $d_\mathrm{m}$ as half the difference between the two maximum excursions on each side centered around the begining point over multiple cycles with an error of $0.06~\mathrm{mm}$.  The shear speed $V_\mathrm{s}$ was calculated as $4d_\mathrm{m}/T_\mathrm{s}$, where $T_\mathrm{s}$ is the period averaged over multiple cycles.  The error of $V_\mathrm{s}$ is less than 3.6\%.

The fingering patterns were imaged using a Prosilica GX 3300 camera from below the bottom plate. We plotted the interface between the two fluids in polar coordinates with $\theta = 0$ ($\theta = \pi$) along the positive (negative) $X$-axis. This is the axis of the shear. To determine the onset radii in different angular directions, we divide the interface into 32 equal segments with the center of the first segment at $\theta = 0$. The onset radius, $R_{\rm on}$, is measured inside each segment. The valleys between fingers are sharper than the fingers themselves; we therefore measure the position of the first minimum with an amplitude of at least $0.2$ mm from its nearest local maximum. The radial coordinate of this first detected minimum in the original interface segment is used as the onset radius for that segment.  To reduce the influence of noise, we fit $R_{\rm on}$ versus $\theta$ with a second-order Fourier component. We used the fitting results as the onset in the parallel directions.

To measure $C(r)$, we dyed the inner fluid with $0.04\%$ (wt) of brilliant blue G-250 ( Alfa Aesar),  and calculated the concentration from calibrated intensity of transmitted light~\cite{bischofberger_fingering_2014,PhysRevFluids.4.033902}. The concentration profile $C(r)$ is tracked along the direction from the center of the pattern to the tip of the interface.  After fingers are merged, the profile is taken across the finger peaks for individual fingers closest to the parallel directions  ($\theta = 0$ and $\theta = \pi$).  Prior to fingering,  the profile is taken along the same azimuthal directions as the earliest-detected finger in subsequent frames.

The finger length $R_{\rm f}$ is tracked for individual fingers close to the parallel directions.  $R_{\rm f}$ is the difference between the radial coordinate of a local maximum and the average of the radial coordinates of its adjacent minima on the interface. 

\textit{Simulations:}  The simulations were done using COMSOL Creeping Flow(spf) and Transport of Concentrated Species (tcs) coupled by Reacting Flow (nirf) under Multiphysics (COMSOL version 6.2.0.278)~\cite{comsol}.
We used a two-dimensional rectangular geometry with gap $b=0.305~\mathrm{mm}$ and length $L=35~ \mathrm{mm}$. The top and bottom walls have no slip boundary conditions with no flux.  
The initial inner fluid concentration was set to an error function $1/2(1-{\rm erf}((x-5)/\delta))$, where $\delta=8.77~\mathrm{\mu m}$ is the initial diffusion width. 
The left wall is the inlet with fully developed flow as the boundary condition and the right wall is the outlet with a static pressure. 

To ensure consistent velocity initial conditions, $U$ is smoothly increased from 0 with a transition zone of $0.01~\mathrm{s}$ and two continuous derivatives.  The top wall was assigned a translational velocity with a square-wave function with $V_\mathrm{s}=12~\mathrm{mm/s}$ as the amplitude and $4d_\mathrm{m}/V_\mathrm{s}=1~\mathrm{s}$ as the period. The phase shift for the square-wave function is $0.25~\mathrm{s}$.  To avoid singularities, the wall motion is smoothed to have a continuous second derivative and smoothly increased from $0~\mathrm{mm/s}$. The viscosities for the inner and outer fluids were set to be $\eta_{\rm in}= 35 ~ \rm{m Pa\cdot s}$ and $\eta_{\rm out}= 218~\rm{mPa\cdot s}$.  The diffusion between the fluids was simulated using Fick's law with diffusion coefficient $D_{12}=1.21\times10^{-10}~\mathrm{m^2/s}$ \cite{PhysRevFluids.4.033902}. We used mapped mesh with rectangular shape as the initial mesh and regular adaptive mesh refinement with rough global minimum around the inner-outer fluid interface to enhance resolution.

\clearpage
\bibliography{apssamp}

\section*{Acknowledgments}
We would like to thank Irmgard Bischofberger, Justin C. Burton, Michelle M Driscoll, Savannah D. Gowen, Mengfei He for insightful discussions.  
This work was primarily supported by the University of Chicago Materials Research Science and Engineering Center, NSF-MRSEC program under award NSF-DMR 2011854.
\paragraph*{Author contributions:}
Preliminary \& exploratory work: TEV, SRN; Conceptualization: SRN, ZL, TEV, SA; Investigation: SRN, ZL; Experiments: ZL; Simulations: ZL; Data analysis and visualization: ZL; Supervision: SRN; Writing-Original draft: ZL; Writing—review \& editing: ZL, SRN, TEV, SA


\subsection*{Supplementary materials}
\noindent
Role of Interface Bluntness on Onset\\
Early-time finger growth rate\\
Concentration profile without shear\\
COMSOL simulations\\
Figs. S1 to S5


\newpage


\renewcommand{\thefigure}{S\arabic{figure}}
\renewcommand{\thetable}{S\arabic{table}}
\renewcommand{\theequation}{S\arabic{equation}}
\renewcommand{\thepage}{S\arabic{page}}
\setcounter{figure}{0}
\setcounter{table}{0}
\setcounter{equation}{0}
\setcounter{page}{1} 

\appendix*
\hypertarget{Sec:SI}{}
\section{SI}
\noindent This PDF file includes:
\begin{itemize}
  \item Role of Interface Bluntness on Onset
  \item Early-time finger growth rate
  \item Concentration profile without shear
  \item COMSOL simulations
  \item Figs. S1 to S5 
\end{itemize}

\subsection{Role of Interface Bluntness on Onset  }
The shape of the interface as a less-viscous miscible fluid invades a more viscous one is determined by several distinct mechanisms.  Aside from the role of shear, as discussed in the present paper, the viscosity ratio of the two fluids, $\eta_\mathrm{in}/\eta_\mathrm{out}$, plays a crucial role. As $\eta_\mathrm{in}/\eta_\mathrm{out}$ increases towards unity, the profile of the inner fluid near its tip becomes smoother and $R_\mathrm{on}$ increases~\cite{bischofberger_fingering_2014}.  That is,  $R_\mathrm{on}$ increases with decreasing $|C'|_{\rm tip}$ which is either due to an increasing  $\eta_\mathrm{in}/\eta_\mathrm{out}$ or to the presence of shear.

Here we compare how these two ways of smoothing the profile (\textit{i.e.}, decreasing $|C'|_{\rm tip}$) affect the onset. We vary $|C'|_{\rm tip}$ by (i) varying the viscosity ratio without any shear or (ii) we keep the viscosity ratio constant and vary the shear amplitude. In both sets of experiments, we use identical Hele-Shaw cell geometries. 


In the case that shear is applied, $|C'|_\mathrm{tip}$ depends only on $d_\mathrm{m}$, but $R_\mathrm{on}$ depends on both $d_\mathrm{m}$ and $V_\mathrm{s}$, in a way that 
increasing $V_\mathrm{s}/Q$ decreases $r_\mathrm{0}$ delaying  $R_\mathrm{on}$  (\hyperlink{fig:r0_Ron}{Fig.~\ref{r0_Ron}a}). However, as shown in \hyperlink{fig:Cf_Ron}{Fig.~\ref{Cf_Ron}b}, when $V_\mathrm{s}/Q$ keeps increasing, $r_\mathrm{0}\rightarrow 0$ and $R_\mathrm{on}$ plateaus. For a sufficiently high shear speed (\textit{e.g.}, $V_\mathrm{s}/Q>0.11~\mathrm{mm^{-2}}$), the measured $R_\mathrm{on}$ is the one corresponding to  $|C'|_\mathrm{tip}=|C'|_\mathrm{final}$. 
The shear amplitudes $d_\mathrm{m}$ vary between $0.3~\mathrm{mm}$ and $1.6~\mathrm{mm}$. 

\begin{figure}
    \hypertarget{fig:together}{}
    \centering
    \renewcommand{\thefigure}{S\arabic{figure}}
    \includegraphics[width=\linewidth]{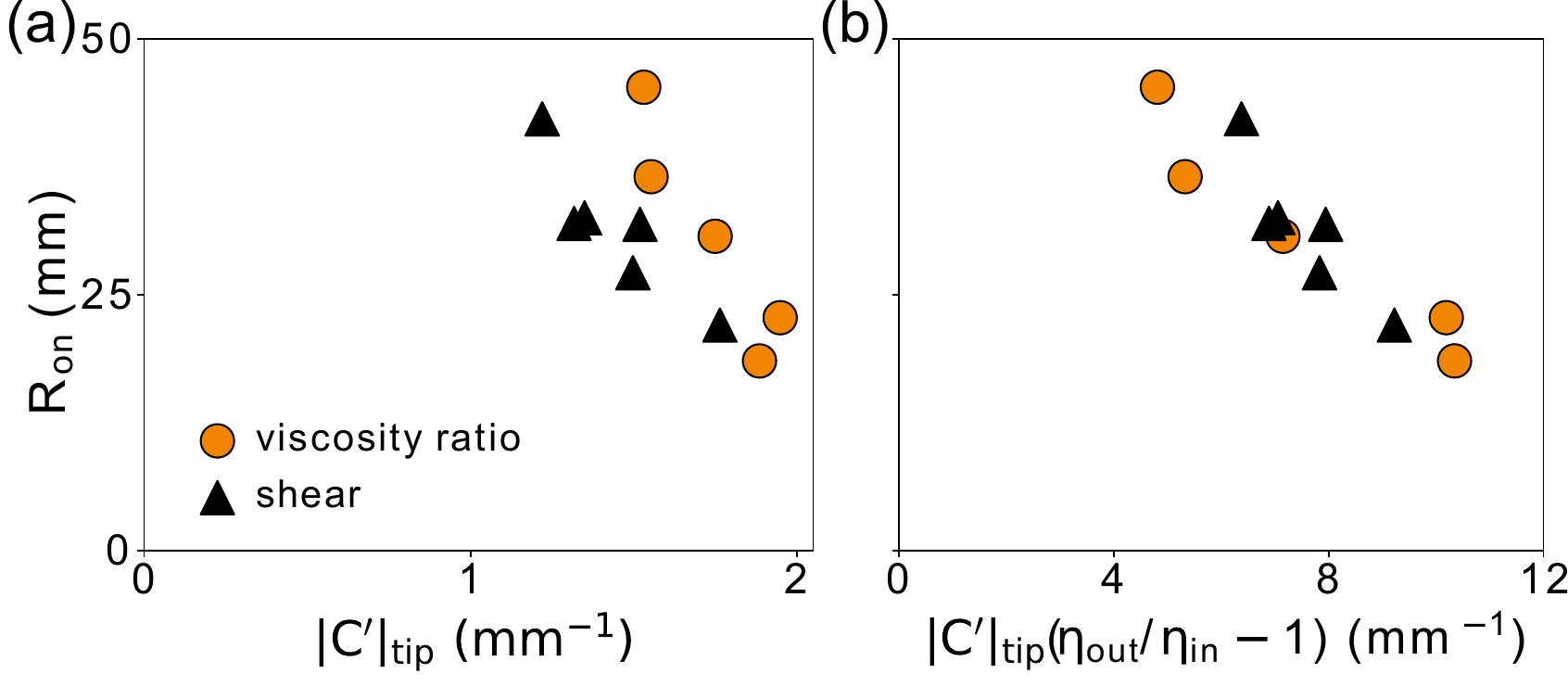}
\caption{\textbf{Comparison of onset radius versus profile shape for two distinct control parameters.} \textbf{(a)} Onset radius, $R_{\mathrm{on}}$, versus bluntness of interface tip, $|C'|_{\mathrm{tip}}$. Black triangles: at fixed viscosity ratio ($\eta_{in}/\eta_{out} = 0.16$ as in manuscript) varying shear ($0.3~\mathrm{mm} < d_\mathrm{m} < 1.6~\mathrm{mm}$) with $V_\mathrm{s}/Q>0.11 ~\mathrm{mm^{-2}}$. 
Oranges circles: no shear but varying viscosity ratio ($0.15 < \eta_{in}/\eta_{out} < 0.24$). Both data sets have the same overall trend. 
\textbf{(b)} Including a scaling factor of $(\eta_\mathrm{out}/\eta_\mathrm{in} - 1)$ on the abscissa produces a somewhat better collapse of the two two data sets; however, additional data are required to identify a convincing scaling factor.   
}
    \label{together}
\end{figure}

We plot $R_\mathrm{on}$ versus $|C'|_\mathrm{tip}$ for both sets of experiments in \hyperlink{fig:together}{Fig.~\ref{together}}. The data share a similar trend and are close to each other, as shown in \hyperlink{fig:together}{Fig.~\ref{together}a}. We note that if we scale the abscissa by a factor $(\eta_\mathrm{out}/\eta_\mathrm{in} - 1)$ the collapse of the two data sets is slightly better as shown in \hyperlink{fig:together}{Fig.~\ref{together}b}. More data is needed to establish which of these two scaling forms is a better fit. 

\subsection{Early-time finger growth rate}
\begin{figure}
    \hypertarget{fig:Rf_Ro}{}
    \centering
    \renewcommand{\thefigure}{S\arabic{figure}}
    \includegraphics[width=\linewidth]{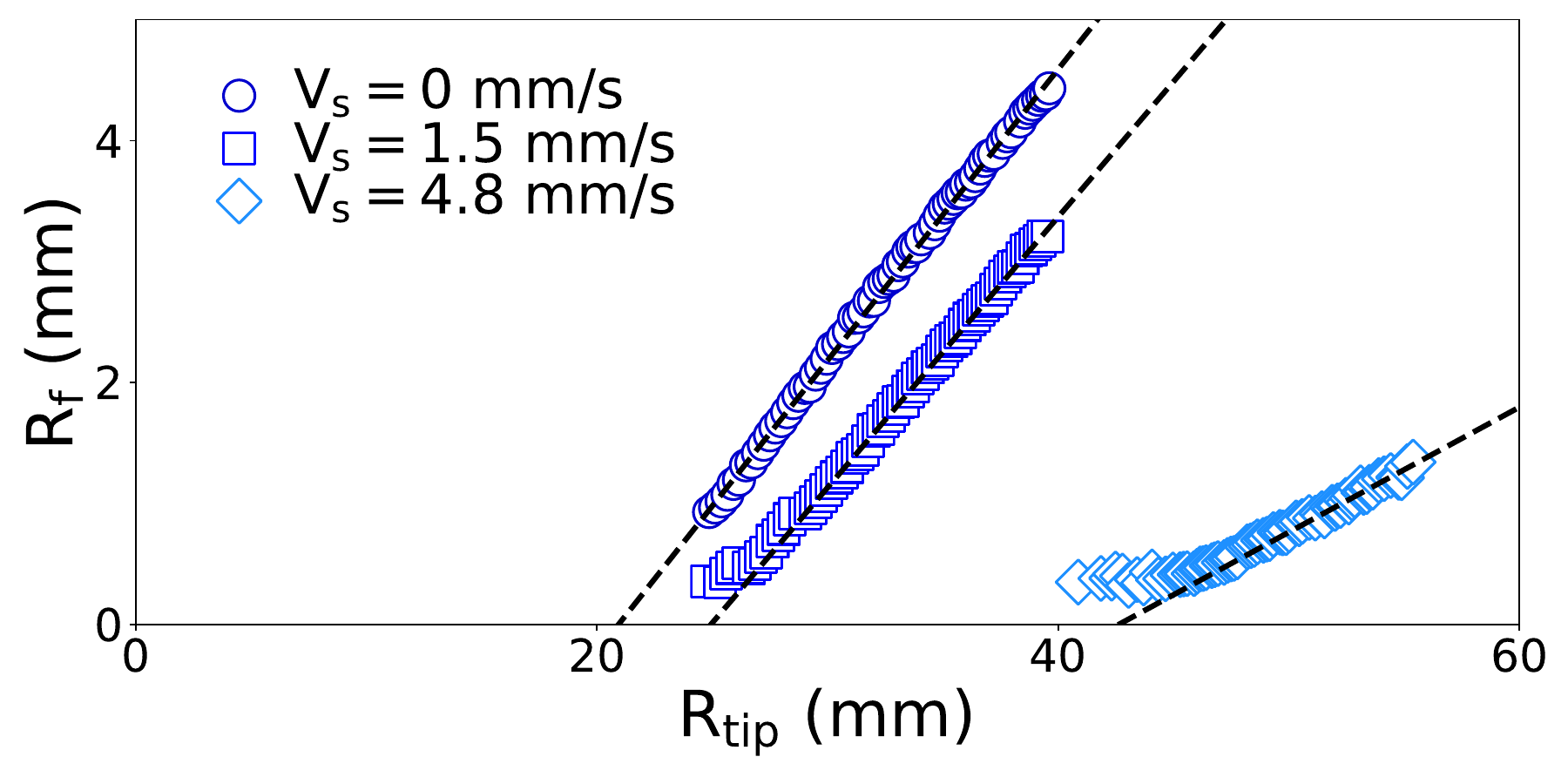}
    \caption{\textbf{Finger length $R_\mathrm{f}$ versus the pattern size $R_\mathrm{tip}$.}  For each dataset, the finger closest to the $\theta=\pi$ direction is tracked after $R_\mathrm{f}$ exceeds 0.2 mm. Following an initial exponential region, $R_\mathrm{f}$ grows linearly with $R_\mathrm{tip}$, which can be fitted linearly with the dashed lines. The slope of each fit represents the finger growth rate $\Gamma=dR_\mathrm{f}/dR_\mathrm{tip}$, with steeper slopes corresponding to faster growth. For data with shear, $d_\mathrm{m}=3~\mathrm{mm}$ and {$Q=133~\mathrm{\mu L/s}$}.    
    }
    \label{Rf_Ro}
\end{figure}

At a constant injection rate,  the interface propagation speed $U$ decreases as the pattern size increases, which in turn reduces $dR_\mathrm{f}/dt$ . To account for this effect, we evaluate the finger growth rate relative to the interface growth rate by examining the ratio  $\frac{dR_\mathrm{f}/dt}{dR_\mathrm{tip}/dt}=dR_\mathrm{f}/dR_\mathrm{tip}$. We know that when there is no shear, finger length initially grows exponentially with pattern size and then transitions to a linear regime,  where $\Gamma=dR_\mathrm{f}/dR_\mathrm{tip}$ is a constant~\cite{videbaek2020delayed}  This linear growth behavior persists under shear across different shear speeds, as shown in \hyperlink{fig:concentrationprofile}{Fig.~\ref{Rf_Ro}}. We measure the finger growth rate by fitting the linear regime of each curve and using the slope of the fit as the value of $\Gamma$.

\subsection{Concentration profile without shear}

\begin{figure}
    \hypertarget{fig:c_noshear}{}
    \centering
    \renewcommand{\thefigure}{S\arabic{figure}}
    \includegraphics[width=\linewidth]{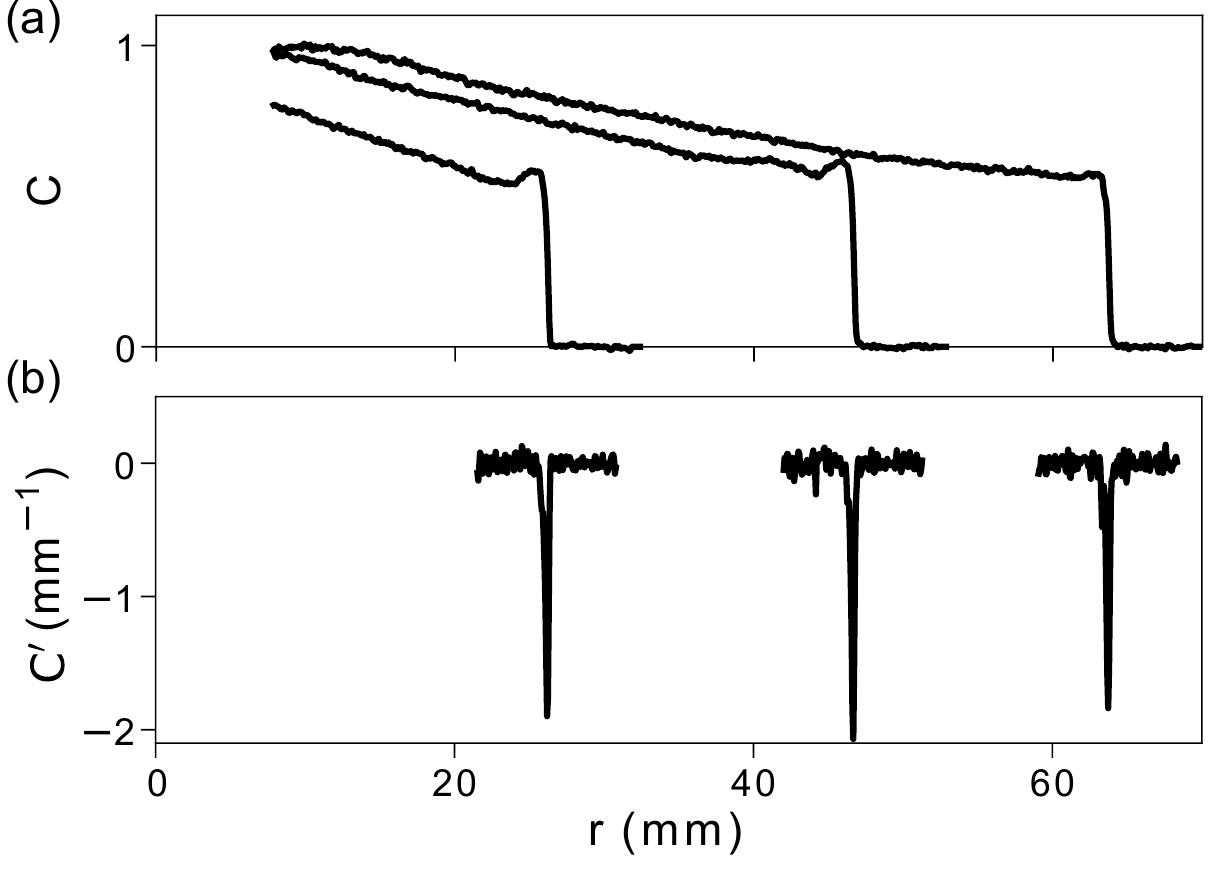}
\caption{\textbf{Comparison of concentration profiles with and without shear} 
\textbf{(a)} Evolution of inner fluid concentration profiles $C(r)$ from an experiment without shear as the interface propagates outwards. The parameters for the Hele-Shaw cell geometry and the injection rate are the same as the case in \protect\hyperlink{fig:concentrationprofile}{Fig.~\ref{concentrationprofile}a} wtih shear. The curves from left to right correspond to the profiles at $R_\mathrm{tip}\approx 26~\mathrm{mm}, 47~\mathrm{mm}, \mathrm{and}~64~\mathrm{mm}.$ \textbf{(b)} Radial derivatives of concentration profiles $C'(r)$ near the tip, corresponding to the curves in \textbf{(a)}.}
   \label{fig:c_noshear}
\end{figure}

In \hyperlink{fig:c_noshear}{Fig.~\ref{fig:c_noshear}a}, we plot the evolution of the concentration profiles without shear for a direct comparison with the profiles when there is shear shown in \hyperlink{fig:concentrationprofile}{Fig.~\ref{concentrationprofile}a}. Both experiments use the same Hele-Shaw cell geometry, parameters, and fluids. In the absence of shear, the interface tip remains blunt as the inner fluid expands outwards. This is shown in \hyperlink{fig:concentrationprofile}{Fig.~\ref{concentrationprofile}} and more obviously in \hyperlink{fig:c_noshear}{Fig.~\ref{fig:c_noshear}b}, where $|C'|_\mathrm{tip}$ only fluctuates but does not exhibit a sustained decrease with increasing radius as the inner fluid expands outwards. 

\subsection{COMSOL simulations}
\begin{figure}[htbp]
    \hypertarget{fig:onecycle}{}
    \centering
    \renewcommand{\thefigure}{S\arabic{figure}}
    \includegraphics[width=\linewidth]{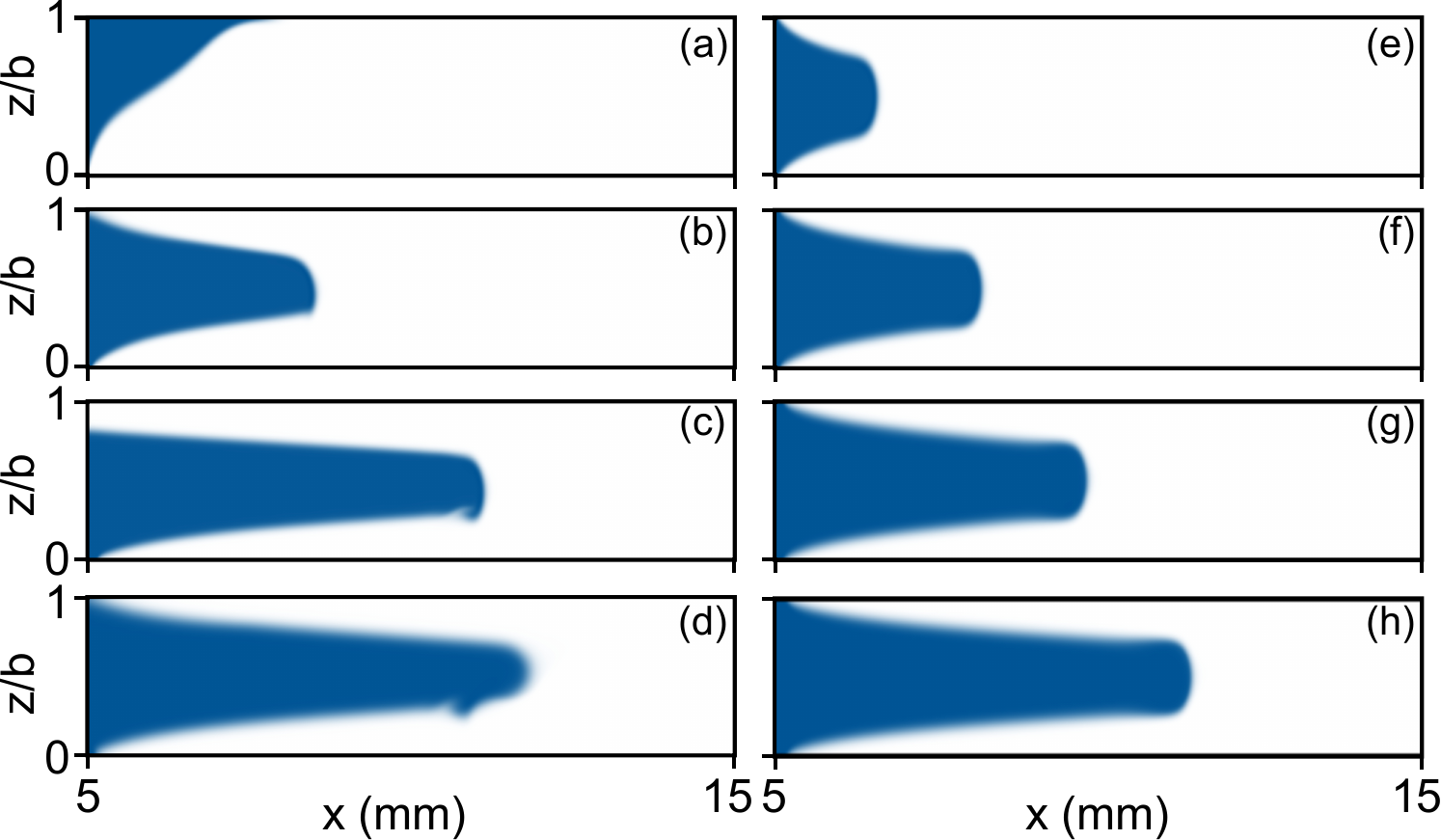}
    \caption{\textbf{Simulation showing shape of inner fluid during the first shear cycle.}
    \textbf{(a-d)} COMSOL simulation with $V_\mathrm{s}=12~\mathrm{mm/s}$, $U=4~\mathrm{mm/s}$, and $d_\mathrm{m}=3~\mathrm{mm}$ during the first shear cycle:
    \textbf{(a)} top plate at the right-most side;
    \textbf{(b)} top plate back to center after half a cycle;
    \textbf{(c)} top plate at the left-most side;
    \textbf{(d)} top plate back to center after one full cycle.
    \textbf{(e-h)} Simulations without shear at the same times as \textbf{(a-d)}, using the same fluid properties, geometry, and interface velocity as the shear case.
    Even when the plate is moving opposite to the interface, the tip under shear is still smoother and narrower.
    Simulations start with the inner fluid initially filled to $5~\mathrm{mm}$.
    Inner fluid is blue, outer fluid is white.}
    \label{fig:onecycle}
\end{figure}

\begin{figure}[htbp]
    \hypertarget{fig:stable}{}
    \centering
    \renewcommand{\thefigure}{S\arabic{figure}}
    \includegraphics[width=\linewidth]{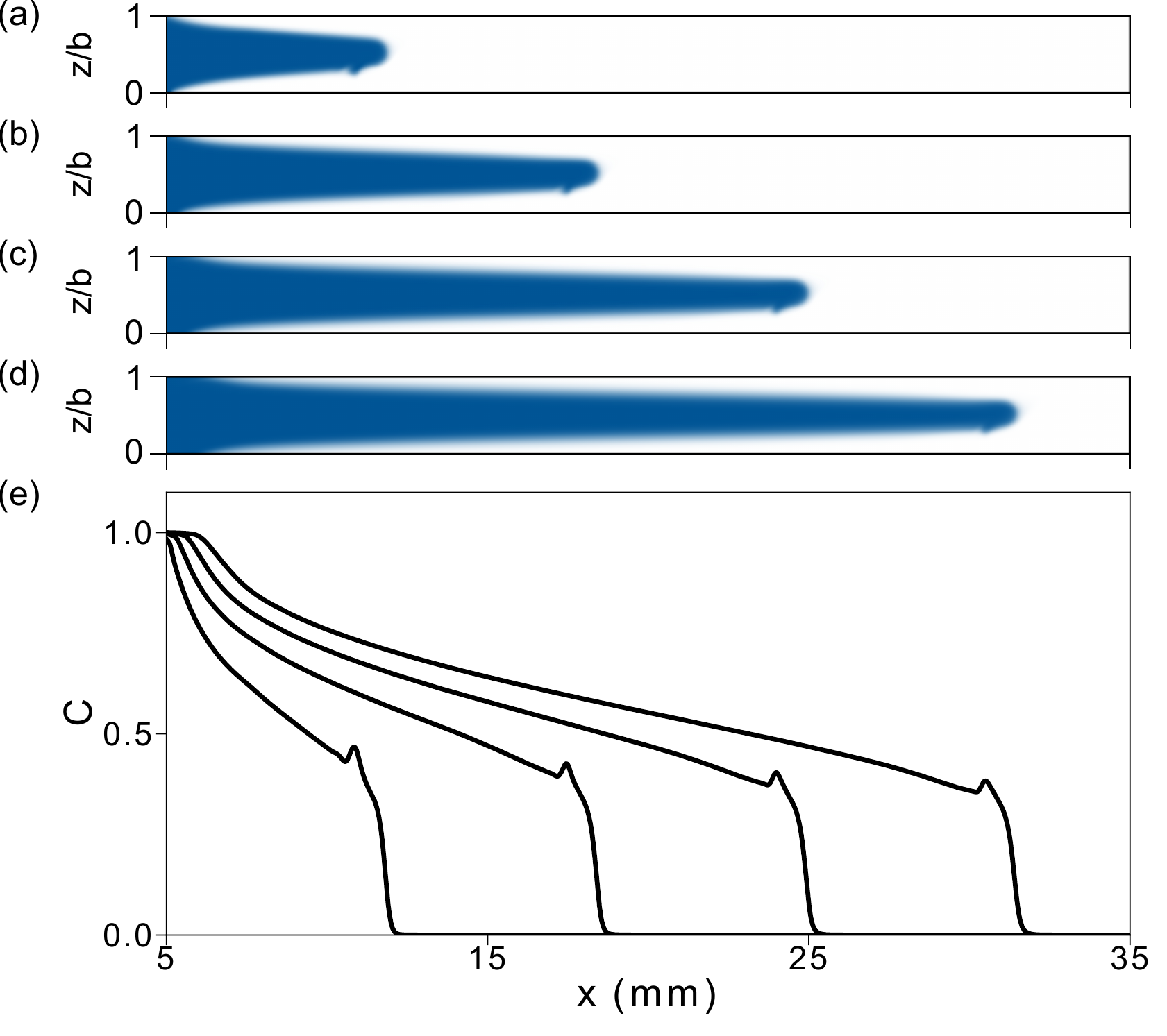}
\caption{\textbf{Simulation of the inner fluid interface after multiple cycles.} \textbf{(a-d)} COMSOL simulation shows the shape of the inner fluid after one (a), two (b), three (c), and four (d) cycles.  \textbf{(e)} The concentration profiles corresponding to the interfaces in (a-d) from left to right. The overall shape of the profile remains the same although the width of the profile becomes slightly more slender at later times.}
    \label{fig:stable}
\end{figure}

We use COMSOL simulations to investigate how shear perturbs the shape of the interface during the first shear cycle as shown in \hyperlink{fig:onecycle}{Fig.~\ref{fig:onecycle}}. After the top plate moves to the right during the first quarter of a cycle, the interface is dragged to be significantly tilted (\hyperlink{fig:onecycle}{Fig.~\ref{fig:onecycle}a}). When the top plate reverses direction to the left, the interface becomes sharper again (\hyperlink{fig:onecycle}{Fig.~\ref{fig:onecycle}b,c}), but still smoother and narrower compared with the no shear case  (\hyperlink{fig:onecycle}{Fig.~\ref{fig:onecycle}f,g}). 
To investigate the long-time stability of the interface, we simulate the profile at the end of subsequent cycles (\hyperlink{fig:stable}{Fig.~\ref{fig:stable}a-d}).  After multiple cycles, the interface becomes slightly thinner as shown in (\hyperlink{fig:stable}{Fig.~\ref{fig:stable}e}) but the interface retains the overall shape of (\hyperlink{fig:onecycle}{Fig.~\ref{fig:onecycle}d}).   

\end{document}